\input{aipcheck}

\documentclass[
    ,final            
  ]
  {aipproc}

\layoutstyle{6x9}

\def\eps{\epsilon}
\def\epse{\eps_{e,-1}}

\def\cm3{\mbox{cm}^{-3}}

\def\tdec2{t_{dec,2}}
\def\cm{\mbox{cm}}
\def\ergscm{\mbox{ergs s}^{-1}\mbox{cm}^{-2}}
\begin{document}

\title{High-energy gamma-rays from GRB X-ray flares }

\classification{}
\keywords     {}

\author{X. Y. Wang}{
  address={Department of Astronomy and
Astrophysics, Penn. State University, University Park, PA
16802},altaddress={Department of Astronomy, Nanjing University,
Nanjing 210093, China }}

\author{Z. Li}{
  address={Physics Faculty, Weizmann Institute of Science, Rehovot
76100, Israel } } \iftrue
\author{P. M\'esz\'aros}{
  address={Department of Astronomy and
Astrophysics, Penn. State University, University Park, PA 16802}  } 

\begin{abstract}
 The recent detection of X-ray flares during the afterglow
phase of gamma-ray bursts (GRBs) suggests an inner-engine origin,
at radii inside the forward shock. There must be inverse Compton
(IC) emission arising from such flare photons scattered by forward
shock afterglow electrons when they are passing through the
forward shock. We find that this IC emission produces   high
energy gamma-ray  flares, which may be detected by AGILE, GLAST
and ground-based TeV telescopes. The anisotropic IC scattering
between flare photons and forward shock electrons does not affect
the total IC component intensity, but cause a time delay of the IC
component peak relative to the flare peak. The anisotropic
scattering effect may also weaken, to some extent, the suppression
effect of the afterglow intensity induced by the enhanced electron
cooling due to flare photons. We speculate that this IC component
may already have been detected by EGRET from a very strong
burst---GRB940217. Future observations by GLAST may help to
distinguish whether X-ray flares originate from late central
engine activity or from external shocks.
\end{abstract}

\maketitle


\section{IC of  flare photons by forward shock electrons}
X-ray flares have been detected during the early afterglows in a
significant fraction of gamma-ray bursts (e.g. \cite{Burrows05};
\cite{Zhang06}; \cite{O'Brien06}). The amplitude of the X-ray
flare can be a factor of up to $\sim 500$ for GRB050502B
\cite{Burrows05, Falcone06}, and in most cases a factor several,
compared with the background afterglow component. The rapid rise
and decay behavior of some flares suggests that they are caused by
internal dissipation of energy due to late central engine
activity. Therefore the inner flare photons, when they are passing
through the forward shocks, must interact with the shocked
electrons and get boosted to higher energies \cite{WLM06}. See
Fig.1 for a cartoon illustration. The observed overlapping in time
between flares and afterglows indicates that these two kind of
particles (i.e. photons and shocked electrons) have indeed
interacted with each other  around the overlap
time\cite{Beloborodov05}. This process is expected to occur
whether the X-ray flares are produced by internal shocks or by
other internal energy dissipation mechanisms, as long as they
occur inside the inner edge of the forward shock region.

\begin{figure}
\caption{Cartoon illustration of the IC scattering between flare
photons and forward shock electrons.}
\includegraphics[height=0.25\textheight]{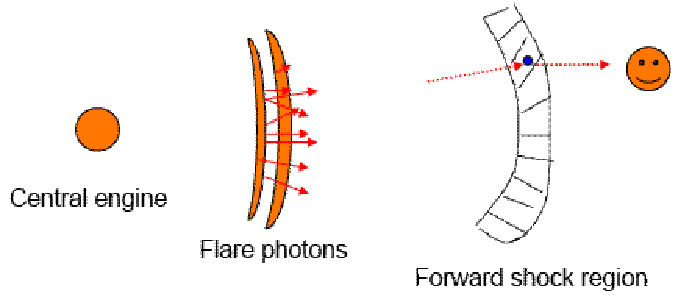}
\end{figure}
Let us now estimate the flux of this IC component. We consider an
X-ray flare of duration $\delta t$ superimposed upon an underlying
power law X-ray afterglow around time $t=10^3t_3 \,{\rm s}$ after
the burst, as observed in GRB050502B. We find that the energy
density of flare photons in the comoving frame of the forward
shock is larger than the magnetic energy density ($B^2/8\pi$) when
$F_X>10^{-10} \epsilon_{B,-2}E_{52}t_3^{-1}D_{28}^{-2} \, {\rm erg
cm^{-2} s^{-1}}$, where $\epsilon_B$ is the equipartition factor
for magnetic field in forward shocks. Since the flux of the X-ray
flare is also much larger than that of the underlying X-ray
afterglow which may represent the synchrotron luminosity, we
conclude that generally the cooling of forward shock electrons  is
dominated by scattering  the X-ray flare photons. It is also shown
that when the X-ray flare flux is larger than a critical flux
$F_{X,c}=3\times10^{-10}E_{52}^{1/2}\epsilon_{e,-1}^{-1}n_0^{-1/2}
t_3^{-1/2}D_{28}^{-2}\ergscm$,  we have $\gamma_m>\gamma_c$, and
all the newly shocked electrons will cool, emitting most of their
energy into  the IC emission\cite{WLM06}. This critical flux is
usually much lower than the X-ray flare flux averaged over its
duration, so we conclude that the flare photons can effectively
cool the electrons in the forward shock and most energy of the
newly shock electrons will be lost into IC emission.  So the
total IC energy is $E_{IC}=\delta t L_e=2\times
10^{51}\epsilon_{e,-1}E_{52}(\delta t/t_p) {\rm erg} $ and the
averaged IC flux is
\begin{equation}
F_{IC}\simeq10^{-9}\epse E_{52}t_{p,3}^{-1}D_{28}^{-2}({\delta
t}/{t_p})\ergscm ,
\end{equation}
where $\delta t< t=t_p$ and $t_p$ denotes the peak time of the
flare \cite{WLM06}. The observed IC $\nu F_\nu$ flux peaks at
\begin{equation}
\varepsilon_{IC,p}\simeq 2\gamma_m^2\varepsilon_X\simeq 3
\epse^2E_{52}^{1/4}n_0^{-1/4}t_3^{-3/4}\varepsilon_{X,\rm keV}
\mbox {GeV}
\end{equation}
where $\varepsilon_X$ is the  peak  of the  flare energy spectrum.
The IC emission occurs right inside the GLAST window and has a
total fluence about $10^{-7}$--$10^{-6}\epse E_{52}D_{28}^{-2}\rm
erg \, cm^{-2}$ for $0.1<\delta t/t<1$, so it should be detected
by GLAST. For the strongest bursts with $E\sim10^{54}\rm erg $,
the GeV photons  could even have been detected by EGRET, such as
from GRB940217 \cite{Hurley94}. AGIlE may also be able to detect
this high energy component from strong bursts.

If described approximately as broken power law, the IC energy
spectrum($\nu F_{\nu}$) has indices of $1/2$ and $-(p-2)/2$ before
and after the break at $\varepsilon_{IC,p}$ respectively. The
$-(p-2)/2$ power law spectrum can extend to a maximum energy
$\varepsilon_{IC, M}$, above which the IC falls into the
Klein-Nishina regime. Requiring the IC scattering to be in the
Thomson regime,$\gamma_e<\gamma_{e,M}= { {\Gamma m_e
c^2}/{\varepsilon_X}}$, gives $\varepsilon_{IC, M}=0.4
E_{52}^{1/4}n_0^{-1/4}t_3^{-3/4}\varepsilon_{X,\rm keV}^{-1}\,{\rm
TeV}$.  The optical depth due to $\gamma\gamma$ absorption on the
X-ray flare photons for the maximum energy $\varepsilon_{IC, M}$
is $\tau_{\gamma\gamma}\simeq
0.3{F_{X,-9}n_0^{1/2}t_3^{1/2}}{E_{52}^{-1/2}}D_{28}^2({\delta
t}/{t})\varepsilon_{X,\rm keV}^{-1}$. Due to the low absorption
depth and the flat spectral slope above $\varepsilon_{IC, M}$, TeV
photons associated with bright X-ray flares are expected to be
detectable with  detectors such as MAGIC, Milagro, H.E.S.S.,
VERITAS and ARGO  etc.

\section{The anisotropic IC scattering effect}
 The incoming X-ray flare photons, which moves out in radial direction, are  anisotropic
seen by the isotropically distributed electrons in the forward
shock, so more head-on scatterings may decrease the IC emission in
the $1/\Gamma$ cone  along the direction of the photon beam
(corresponding to the angle less than $\pi/2$ relative to the
photon beam direction in the comoving frame) \cite{Ghisellini91},
but enhance the emission at larger angles, with  about a fraction
of $3/8$ of total emission falling into the angles between
$1/\Gamma$ and $2/\Gamma$ \cite{WM06}.   In the comoving frame of
the forward shock, the IC emission power depends on the relative
angle $\theta_s$ between the scattered photon direction and the
seed photon beam direction as $P(\theta_s)\propto (1- {\rm
cos}\theta_s)^2$ \cite{RL79}. The factor $(1-{\rm cos}\theta_s)^2$
results in more power emitted at large angles relative to the seed
photon beam direction. By simple algebraical calculation, one can
obtain the fraction of photons scattered into the angles
$0\le\theta_s\le\pi/2$ in the shock comoving  frame (corresponding
to the $1/\Gamma$ cone of the relativistic afterglow jet in the
observer frame due to aberration of light), i.e.
$f_{s}\simeq{\int_0^{\pi/2}(1-{\rm cos}\theta_s)^2{\rm
sin}\theta_s d\theta_s}/{\int_0^{\pi}(1-{\rm cos}\theta_s)^2{\rm
sin}\theta_s d\theta_s}={1}/{8}$. Similarly, one can get that the
fraction of photons falling into the cone of $2/\Gamma$
(corresponding to $\theta_s=2 {\rm arctg} 2$) is $1/2$. So the
effect of anisotropic scattering can decrease the IC emission in
the $1/\Gamma$ cone along the direction of the photon beam, but
enhance the emission at larger angles. However if the opening
angle is $\theta_j\gg1/\Gamma$, the jet geometry can be
approximately regarded  as a sphere, and the observed IC power
after integration over angles  should be the same  in every
direction  in the observer frame \cite{FP06, WLM06}. Supposing
that the flare photons and forward shock ejecta have a $4\pi$
solid angle (i.e. not jet), then the IC emission in the observer
frame should have the same flux in every direction with a flux
level as estimated in the above section. Hence the received IC
component fluence is not reduced.

The anisotropic scattering effect does affect the temporal
behavior of the IC emission. IC emission will be lengthened by the
forward shock angular spreading time, which is also influenced by
the anisotropic IC effect, so the IC component would not be
correlated strictly with the flare light  curves. Since a large
fraction of the IC emission is falling within angles $\theta_{IC}$
between $1/\Gamma$ to $2/\Gamma$, the IC component will peak at
\begin{equation}
t_{IC,p}\sim R/2\Gamma^2 c +R(1-{\rm cos}\theta_{IC})/c\sim
R/2\Gamma^2 c+R\theta_{IC}^2/2c\sim R/\Gamma^2 c\sim 2 t_{{\rm
flare}, p}
\end{equation}
where $t_{{\rm flare}, p}\sim R/2\Gamma^2 c$ is the flare peak
time. The IC component light curves is shown as the thick solid
line in Fig.2. The self-IC of X-ray flares may also produce a
high-energy component, but its light curve should be correlated
with the x-ray flare itself to some extent \cite{WLM06, Galli07},
as shown in the thin solid line. GLAST might be able to
distinguish these two different processes in future with the
temporal behavior of the very high-energy emission and therefore
provides a useful approach for distinguishing whether X-ray flares
originate from late central engine activity or from external
shocks.

\begin{figure}
\caption{A schematic cartoon for the light curves of the X-ray
flares (lower panel) and the predicted IC emission (upper panel)
at GeV-TeV energy  from different processes. In the upper panel,
the solid line represents the IC emission of X-ray flare photons
scattered by afterglow electrons, while the think solid line
denotes the self-IC emission of flare photons in the context of
internal-shock or external-shock scenarios for X-ray flares. }
\includegraphics[height=0.4\textheight]{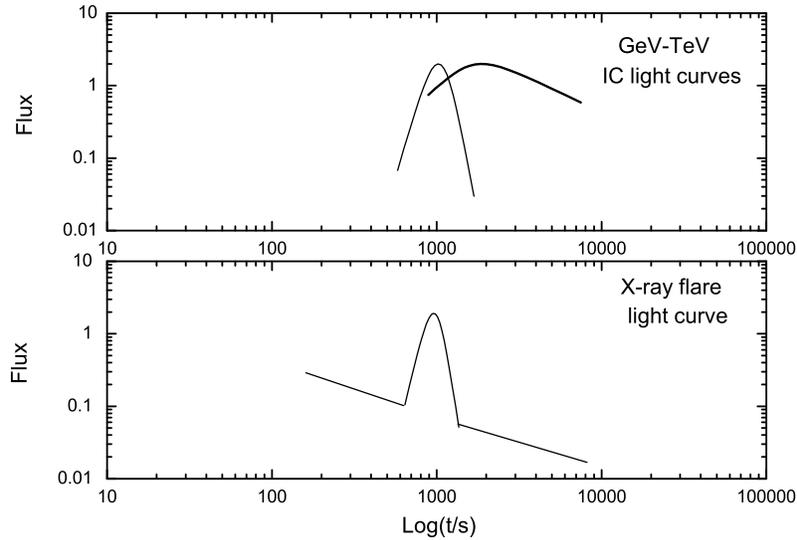}
\end{figure}

The enhanced cooling of the forward shock electrons by the X-ray
flare photons  may suppress the synchrotron emission of the
afterglows during the flare period\cite{WLM06}. Due to more
head-on scatterings with electrons moving in angles larger than
$\pi/2$ relative to the photon beam direction in the comoving
frame and less scatterings with electrons within $\pi/2$, the
anisotropic scattering effect may weaken the cooling effect of
flare photons on the afterglow electrons (which are moving within
$\pi/2$ in the comoving frame and moving within $1/\Gamma$ in the
observer frame correspondingly), hence reduce the extent to which
the flare photons affect the afterglow flux intensity. To suppress
the synchrotron afterglow emission significantly, the cooling
induced by the flare photons should be  higher than that by the
magnetic field.  As a result, only bright flares are expected to
significantly suppress the optical afterglow flux during the flare
period, such as seen in GRB050904 \cite{Gou06}.

\end{document}
\endinput